# FeCAM: A Universal Compact Digital and Analog Content Addressable Memory Using Ferroelectric

Xunzhao Yin, *Member, IEEE*, Chao Li, Qingrong Huang, Li Zhang, Michael Niemier, *Senior Member, IEEE*, Xiaobo Sharon Hu, *Fellow, IEEE*, Cheng Zhuo, *Senior Member, IEEE*, and Kai Ni, *Member, IEEE*

*Abstract*—**Ferroelectric field effect transistors (FeFETs) are being actively investigated with the potential for in-memory computing (IMC) over other nonvolatile memories (NVMs). Content addressable memories (CAMs) are a form of IMC that performs parallel searches for matched entries over a memory array for a given input query. CAMs are widely used for data-centric applications that involve pattern matching and search functionality. To accommodate the ever expanding data, it is attractive to resort to analog CAM for memory density improvement. However, the digital CAM design nowadays based on standard CMOS or emerging NVMs (e.g., resistive storage devices) is already challenging due to area, power, and cost penalties. Thus, it can be extremely expensive to achieve analog CAM with those technologies due to added cell components. As such, we propose, for the first time, a universal compact FeFET-based CAM design, FeCAM, with search and storage functionality enabled in digital and analog domains simultaneously. By exploiting the multilevel-cell (MLC) states of FeFET, FeCAM can store and search inputs in either digital or analog domain. We perform a device-circuit codesign of the proposed FeCAM and validate its functionality and performance using an experimentally calibrated FeFET model. Circuit level simulation results demonstrate that FeCAM can either store continuous matching ranges or encode 3-bit data in a single CAM cell. When compared with the existing digital CMOS-based CAM approaches, FeCAM is found to improve both memory density by 22.4× and energy saving by 8.6×/3.2× for analog/digital modes, respectively. In the CAM-related application, our evaluations show that**

Manuscript received March 20, 2020; revised May 2, 2020 and May 6, 2020; accepted May 9, 2020. This work was supported in part by NSFC under Grant 61974133 and in part by the Applications and Systems-driven Center for Energy-Efficient integrated Nano Technologies (ASCENT), one of six centers in the Joint University Microelectronics Program (JUMP), a Semiconductor Research Corporation (SRC) Program sponsored by Defense Advanced Research Projects Agency (DARPA). The review of this article was arranged by Editor C. Monzio Compagnoni. *(Corresponding authors: Xunzhao Yin; Cheng Zhuo; Kai Ni.)*

Xunzhao Yin, Chao Li, Qingrong Huang, Li Zhang, and Cheng Zhuo are with the College of Information Science and Electronic Engineering, Zhejiang University, Hangzhou 310027, China (e-mail: xzyin1@zju.edu.cn; czhuo@zju.edu.cn).

Michael Niemier and Xiaobo Sharon Hu are with the Department of Computer Science and Engineering, University of Notre Dame, Notre Dame, IN 46556 USA.

Kai Ni is with the Department of Microsystems Engineering, Rochester Institute of Technology, Rochester, NY 14623 USA (e-mail: kai.ni@rit.edu).

Color versions of one or more of the figures in this article are available online at http://ieeexplore.ieee.org.

Digital Object Identifier 10.1109/TED.2020.2994896

**FeCAM can achieve 60.5×/23.1× saving in area/search energy compared with conventional CMOS-based CAMs.**

## I. INTRODUCTION

CONTENT addressable memories (CAMs) are a special form of in-memory computing (IMC) circuits widely used in high-speed searching applications, for example, network routing and CPU caching [1]. As shown in Fig. 1, CAM can compare input query against a list of stored data in parallel, and return the address of matching data or the stored data itself. Thanks to high parallelism and in-memory computation, CAMs have found new utility in emerging deep learning applications [2], [3]. However, there exists two practical challenges preventing the deployment of conventional CMOS static random access memory(SRAM)-based CAMs for deep learning [especially on Internet of things (IoT) devices]: 1) nontrivial power consumption [4] and 2) large area overhead (16 transistors per cell) [5].

To address the aforementioned issues of power and area penalties for SRAM-based CAM, research efforts have been devoted to exploit emerging nonvolatile memories (NVMs) for CAM designs, including resistive RAM (ReRAM) [6], [7], magnetic tunnel junction (MTJ) [8], ferroelectric capacitor (FC) [9] and ferroelectric field effect transistor (FeFET)-based ternary CAM (TCAM) cells [10]–[12]. These NVM-based CAM designs can help reduce area and power consumption while enabling acceleration of various neural network architectures [2], [13]. An FC CAM design was proposed [9] to store 2-bit information in a CAM cell. However, the limited scalability of FC and its read-destructive characteristics are a significant barrier for the wide adoption of the FC CAM [14]. MTJ-based CAM designs, for example, 9T-2MTJ CAM cell [15], mitigate the memory density bottleneck, but the small $R_{ON}/R_{OFF}$ ratio and large write power of MTJs significantly degrade the CAM performance. The CAM designs based on resistive memory devices, including ReRAM [7] and phase change memory (PCM) [6], are advantageous in memory density but have limited $R_{ON}/R_{OFF}$ ratio and significant write power, which is challenging to overcome. In addition, the design also incurs additional complexity and cost due to the necessity of connecting transistors to the back-end-of-line storage elements. Given those challenges, we have designed an ultracompact 2FeFET-based digital TCAM cell by exploiting the large $R_{ON}/R_{OFF}$ ratio, high $R_{OFF}$, three terminal device structure and highly energy-efficient electric field-driven write mechanism of FeFETs [2], [12]. Later on, a similar TCAM cell with different operation principles (e.g., the search lines (SLs) are applied on the FeFET drains; whereas in our TCAM design the SLs are performed on the FeFET gates) is also demonstrated [10]. Therefore, ferroelectric TCAM is highly promising as the most competitive binary/ternary[1] digital CAM candidate [12], [16].

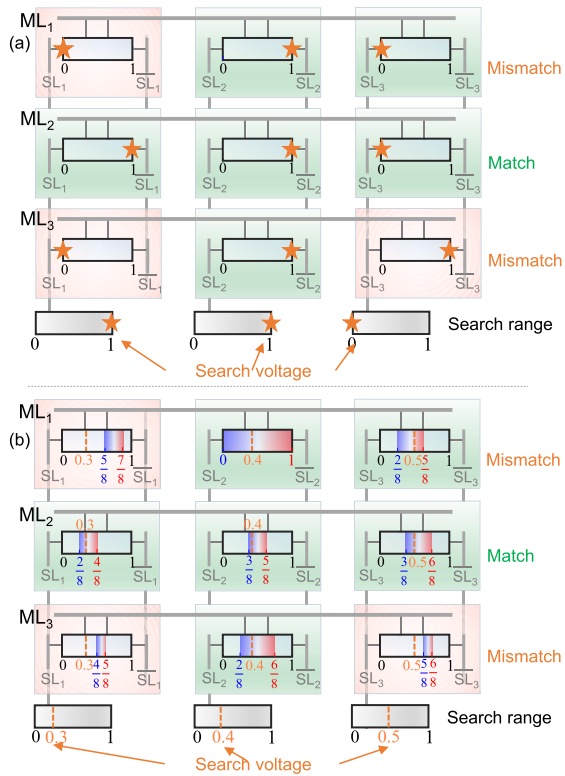

Fig. 1. Examples of (a) digital CAM operation. (b) Analog CAM operation. Matching (mismatching) cells are in green (red).

However, most of the proposed CAM designs to date only support digital storage and search functionality [6], [7], [9]–[12], [15], [17], limiting the CAM density and its functionality. As illustrated in Fig. 1(a), only binary or ternary values are stored and searched in a digital CAM, where the matching cells are colored in green and mismatching cells in red. To avoid the aforementioned limitations, there is a strong need to go beyond binary/TCAM by enabling analog search and storage functionality for CAMs. An analog CAM can store and search analog values within a continuous range, as shown in Fig. 1(b). By exploiting the multilevel-cell (MLC) states in NVMs, such as ReRAM, an analog CAM stores quantized upper and lower bounds of a continuous range, which defines the continuous interval for searching. Therefore, multiple bits can be encoded as the number of nonoverlapping continuous ranges for searching [e.g., 3 bits per cell means eight nonoverlapping continuous ranges, as shown in Fig. 1(b)], thus improving the memory density compared with its binary/ternary digital CAM counterparts and expanding the functionality. Highly promising as it is, designing analog CAMs can be challenging especially when the following characteristics are desired.

[1]Ternary states refer to logic "1," "0," and wildcard "do not care."

1) *Universal:* In near-term applications, digital CAMs, rather than analog CAMs, are dominating CAM-based applications. However, to accommodate the data explosion for future scenarios, it is essential to have the analog CAM design that provides scalable direct analog processing capability for energy and area-efficient processing. The analog CAM design based on ReRAM [18] adopts a different structure from its digital counterpart, thus incurring significant additional cost when both digital and analog specialized CAM designs are required. Therefore, the integration of both digital and analog modes can be complex and costly, if not impossible, in implementation. An efficient and universal CAM implementation has never been created so far.

2) *Compactness:* To achieve high CAM density, minimum (ideally none) transistor overhead is desired to augment a single digital CAM cell to achieve analog functionalities, especially for two-terminal NVM devices. For example, ReRAM-based analog CAM design [18] employs six transistors and two ReRAMs for just one analog CAM cell, consuming a significant area overhead.

Thus, it is highly desirable to have a universal and compact CAM design solution that addresses the aforementioned challenges. In this article, we propose, for the first time, an FeFET-based universal and compact CAM design, FeCAM, which can simultaneously serve as a digital and an analog CAM without any area overhead. Utilizing an experimentally calibrated FeFET model [19], we have performed device-circuit codesign approach of FeCAM. We show that by leveraging the MLC states in FeFETs, a 2FeFET-based CAM design with analog search and storage capabilities is possible. By integrating the proposed analog CAM design with the digital CAM design in [12], our proposed analog CAM design can be expanded to a universal CAM design. As a result, our proposed universal and compact CAM is a highly competitive candidate for associative memory, allowing denser memory density (60.5×), more energy efficiency (23.1×), and flexible digital/analog processing in CAM-related applications compared with conventional CMOS-based CAMs.

## II. ANALOG STATES IN FeFET

The recent discovery of ferroelectric HfO$_2$ has spurred intense research activities in designing CMOS-compatible and high density FeFETs for nonvolatile memory applications [21]. The device operates by applying positive/negative gate pulses to set the ferroelectric polarization direction pointing toward the channel/gate metal direction, setting the FeFET to the low-$V_{TH}$ and high-$V_{TH}$ state, respectively. Unlike other types of NVM devices requiring a large dc conduction current for memory write, a FeFET exhibits superior write energy efficiency since it relies only on the electric field to switch the polarization. Though most of the current research efforts have been focusing on the binary memory property [16], [22], ferroelectric MLC has been studied to increase the memory density [23], [24]. Further device optimization and innovation in the future are likely to be conducted to further boost FeFET memory performance.

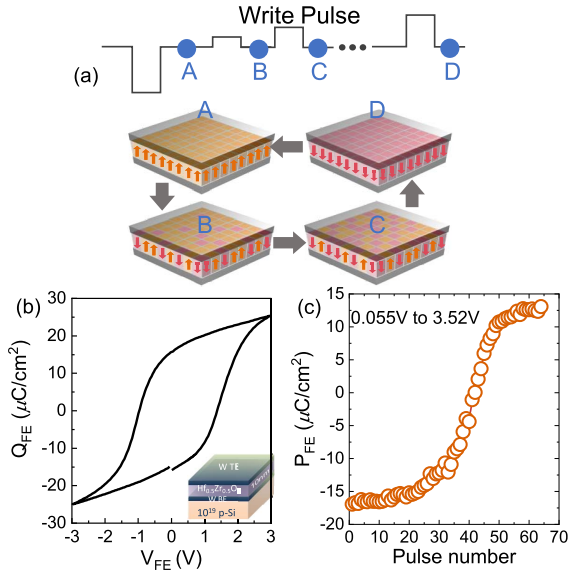

Fig. 2. (a) Schematic of partial polarization switching in an MFM capacitor induced by write pulses with increasing pulse amplitudes. (b) Measured $Q_{FE}$–$V_{FE}$ characteristics from fabricated 10-nm-thick $Hf_{0.5}Zr_{0.5}O_2$ MFM capacitor. (c) Measured polarization as a function of applied pulse number. The detailed device fabrication is presented in [20].

The intermediate $V_{TH}$ states between the low-$V_{TH}$ and high-$V_{TH}$ states have also been utilized for the design of synaptic weight cell in neural network accelerators [25], [26]. The intermediate states are obtained through partial polarization switching, which can be induced by varying applied pulse amplitude or pulsewidth, as illustrated in the metal-ferroelectric-metal (MFM) capacitor shown in Fig. 2(a). Ferroelectric $HfO_2$ thin film is composed of multiple domains with a distribution of their coercive field [20]. Different pulse amplitudes sample different portions of that distribution, inducing partial polarization switching. Fig. 2(c) shows the measured intermediate polarization states in a 10 nm $Hf_{0.5}Zr_{0.5}O_2$ MFM capacitor, whose $Q_{FE}$ − $V_{FE}$ hysteresis loop is shown in Fig. 2(b), under pulse trains with increasing amplitudes. The MFM device details are presented in [20]. These intermediate states lead to different $V_{TH}$s in a FeFET, as shown in Fig. 3.

The experimentally measured $I_D$ − $V_G$ transfer characteristics of a FeFET written with pulses of increasing positive amplitudes are shown in Fig. 3(b). The FeFET details are presented in [21]. The device $V_{TH}$ is gradually reduced with increasing pulse amplitudes. Fig. 3(c) shows the simulated transfer characteristics using a calibrated FeFET compact model [19], which qualitatively reproduces the experimentally observed FeFET analog properties. This model is used for the demonstration of the FeCAM in this article. With the demonstrated analog states in a FeFET, the FeFET-based CAM can provide a viable solution to the analog CAM design. We describe below our proposed FeFET-based universal and compact CAM design integrating both analog and digital modes.

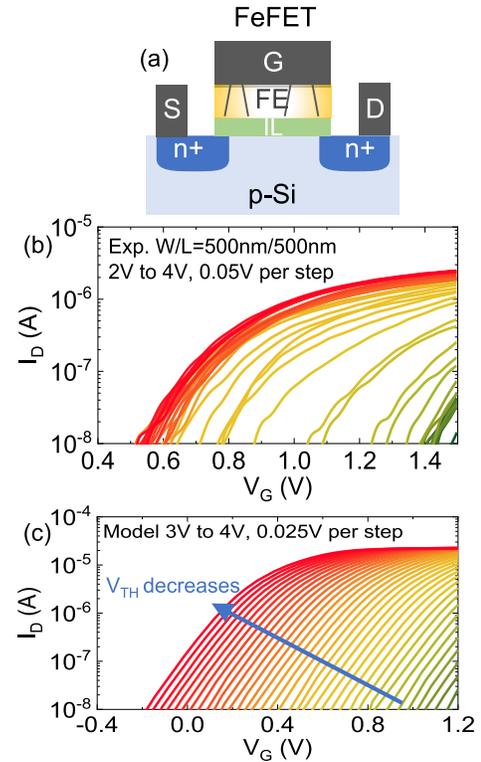

Fig. 3. (a) Schematic of a FeFET device. (b) Measured $I_D$–$V_G$ characteristics from fabricated 10-nm-thick $Hf_{0.5}Zr_{0.5}O_2$ MFM capacitor. (c) Simulated $I_D$–$V_G$ characteristics under increasing write pulse amplitudes. Each curve in (b) and (c) represents the FeFET state after a write pulse with amplitude in the range indicated in the corresponding figures [e.g., from 2 to 4 V with a step of 50 mV in (b)].

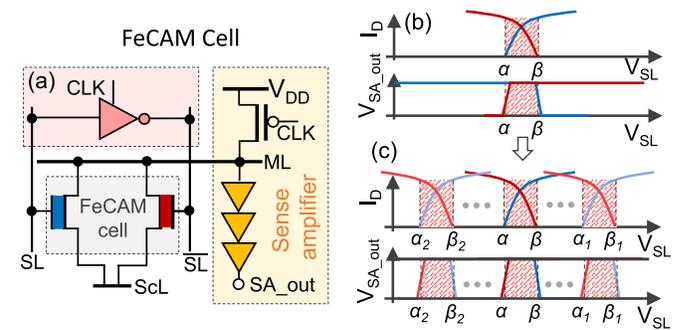

Fig. 4. (a) Proposed universal FeCAM cell and its peripheral circuits, including a clocked inverter for the SL and an SA for the ML. Conceptual illustration of the operations of the CAM cell for (b) matching with only a single voltage range, i.e., a single upper/lower bound. (c) Matching with multiple voltage ranges, i.e., multiple discrete upper/lower bounds. These bounds are determined by the two FeFETs $V_{TH}$. The blue/red curves correspond to the characteristics of the blue/red FeFET in (a), respectively.

## III. FeCAM DESIGN AND OPERATION

### A. Operating Principles of FeCAM

FeCAM is universal in that it can simultaneously function as digital and analog CAM. Fig. 4(a) shows the schematic of the proposed FeCAM cell, which consists of only two FeFETs. The gates of the FeFETs connect to the SL and inverted $\overline{SL}$. The $\overline{SL}$ is generated from the SL using a clocked inverter [Fig. 4(a)]. Since the input of the FeCAM cell is an analog value, an analog inverter with a large transition window in the transfer characteristic is desired. The clock-gated pMOS in Fig. 4(a) is used to precharge the ML before the search operation. A sense amplifier (SA) consisting of three buffers is applied. Fig. 4(b) illustrates conceptually search and storage operations.

Since the digital CAM operations have been discussed in prior works [2], [12], here we focus only on the analog operations of CAM. Each FeCAM cell can store a continuous range of values, which is defined by an upper bound and a lower bound, for matching against the input voltage $V_{SL}$ as shown in Fig. 4(b). When the $V_{SL}$ is smaller than the range upper bound, defined by the blue FeFET [Fig. 4(a)], the blue FeFET is turned off, causing negligible discharge current from ML and hence, leaving ML high. When $V_{SL}$ is larger than the upper bound, the blue FeFET turns on, discharging ML and hence leaving ML low. Due to symmetry, the same behavior can be observed for the red FeFET, with respect to its own gate voltage, $V_{\overline{SL}}$. When plotted as a function of $V_{SL}$, however, its characteristic is flipped horizontally, forming the lower bound for the voltage range. The interval between the two bounds represents the stored voltage range for search. When $V_{SL}$ is between the lower and upper bounds, neither of the device is turned on, resulting in a match output for a range of $V_{SL}$ values. When $V_{SL}$ is outside the defined bound, one of the two FeFETs turns on and discharges ML. As a result, the cell keeps the SA output at high level only when the input $V_{SL}$ falls between the bounds. By properly adjusting the $V_{TH}$ of the FeFETs using the partial polarization switching in Section II, thus shifting the upper and lower bounds, FeCAM allows continuous range storage and searching in multiple bounded regions, as shown in Fig. 4(c). Since the lower and upper bounds of the CAM cell can be independently configured by programming the $V_{TH}$ of the two FeFETs, the CAM cell can be configured, according to the FeFET characteristics (Fig. 3), to create multiple matching voltage ranges. Each range is defined by an upper and lower bound, and the voltage ranges can be adjacent or nonadjacent. As a result, the number of $V_{TH}$ levels corresponds to the number of discrete upper/lower bounds, thus enabling multibit quantized range searching (e.g., eight discrete upper/lower bounds correspond to 3 bits quantized range searching).

### B. FeCAM Cell Characteristics

Fig. 5(a) demonstrates the transient waveforms obtained from SPICE simulations for the search operation in a FeCAM cell. The SA output at three different input voltages ($V_{SL} = 0.3, 0.5, 0.7$ V) are shown, corresponding to below the lower bound, within the bound, and above the upper bound, respectively. The search operation of the FeCAM cell starts after the precharge phase, where the clock signal CLK is low. The inverter that drives the inverted $\overline{SL}$ is powered by CLK. When CLK transits to high, the circuit starts to perform the search operation based on $V_{SL}$. From Fig. 5(a), after a search operation begins, the SA output stays high when $V_{SL}$ is at 0.5 V (within the bound), indicating a match, but falls to low when $V_{SL}$ is at either 0.3 V (below the lower bound) or 0.7 V (above the upper bound) for a mismatch. The transient SA outputs for $V_{SL}$ across the entire voltage range is shown in Fig. 5(b). It suggests that the search result (i.e., whether the search voltage is within the astored range or not) can be measured from the transient SA output at a certain timepoint following the search, where the voltage difference (i.e., sensing margin) between a match and a mismatch is large enough for sensing. The simulated SA outputs at 10 ns following the search operation for different $V_{SL}$ is shown in Fig. 5(c), indicating that the example FeCAM cell stores a continuous voltage matching range of 0.4 and 0.6 V. Fig. 5(d) demonstrates eight ranges, corresponding to 3-bit discrete upper/lower bound levels by independently configuring the $V_{TH}$ of the two FeFETs. This multibit storage can greatly improve the FeCAM information density at a minimal hardware cost. The operating principle and simulations of FeCAM presented above clearly demonstrate that the proposed FeCAM cell implements the desired analog search functionality and can be used as both a digital and analog CAM.

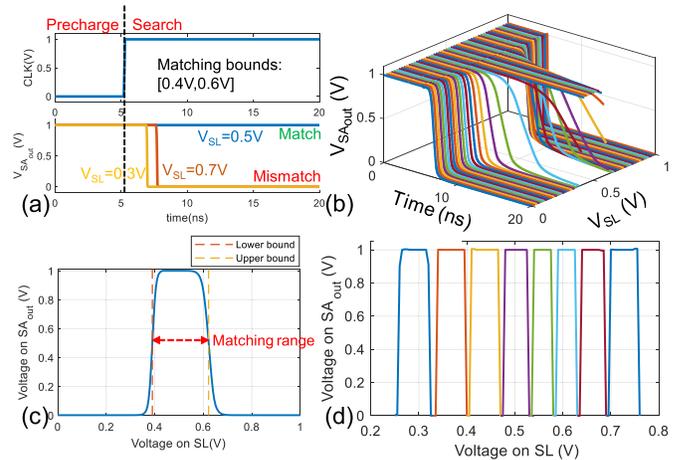

Fig. 5. Search operation of FeCAM. (a) Transient waveform of an FeCAM cell with predefined bounds, which shows a precharge phase (CLK is low) and search phase (CLK is high). (b) 3-D plot for the transient SA output versus different voltages on SL during the search. (c) Corresponding SA output along with different voltages on SL at a timepoint of 10 ns, exhibiting a continuous range for matching. (d) Multiple bounded nonoverlapping continuous range for matching can be realized by adjusting the two FeFETs $V_{TH}$ states in a FeCAM cell. In this article, 3-bit quantized bounds (i.e., eight different upper/lower bounds) are demonstrated.

### C. FeCAM Array Characteristics

The design of a single FeCAM cell and its operation in Sections III-A and III-B clearly demonstrate the capability of multibit storage and search at the cell level. However, when extending from the cell to the array level, the parasitics and accumulated signal deterioration may actually prevent distinguishing different states, causing search failure at the array level. In this section, we provide design guidelines to ensure correct search and storage functionality at the array level without incurring additional design cost. Before presenting the details, we first discuss the multibit search behavior of FeCAM in large arrays.

We simulated the FeCAM arrays based on the array architecture shown in Fig. 6(a). We vary the number of rows and columns to investigate the impact of array size on the multibit search and storage functionality. The wiring parasitics at the FeCAM array are extracted from DESTINY [27]. As shown in Fig. 6(a), the proposed FeCAM array cells can operate in both digital and analog modes simultaneously even at the granularity of single cell level, depending on the write schemes applied. This is quite different from the previous CAMs that can only be either digital or analog [6]–[8], [11], [12], [15], [18], where only one type of value is stored and searched. Such universal operating mode of FeCAM may enable efficient data analytic applications where both exact search (digital mode) and approximate search (analog mode) functions are desired.

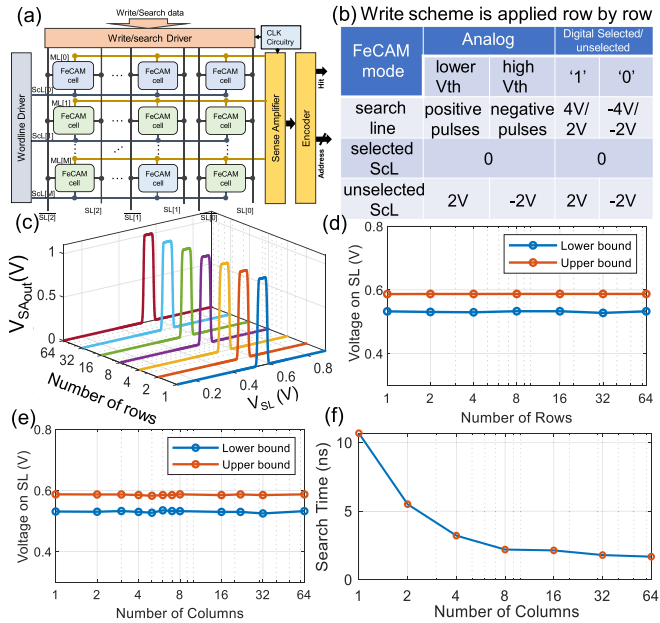

Fig. 6. Simulation of FeCAM array with different sizes. (a) Architecture of the proposed FeCAM array. (b) Write scheme for FeCAM array in analog and digital mode, respectively. The write in the array is performed one row at a time. (c) SA output voltage versus input voltage with different number of rows. (d) Matching range bounds versus the number of rows. (e) Matching range bounds versus the number of columns. (f) Corresponding search time of FeCAM versus the number of columns.

Fig. 6(b) summarizes the write operations of the FeCAM array. In addition to the write scheme for the digital mode of FeCAM similar to [12], we propose the write scheme for the analog mode of FeCAM array. It is based on the inhibition bias schemes, $V_W/2$, presented in [28]. The write for FeCAM array is conducted row-wise, namely one row is written at a time. For the selected row to be written, write pulses are applied to the associated SL and $\overline{\text{SL}}$ according to the FeFET characteristics shown in Fig. 3(c), and the associated source lines, ScLs, are grounded, so that the corresponding write voltage values are applied to the targeted devices. For the unselected rows, the associated ScLs are set to 2 or $-2$ V depending on the analog state to be written to the cell in the selected row and the same column. Since the max voltage of the write pulses is 4 V, the gate–source voltages of the FeFETs in unselected cells should not exceed 2 V. It has been shown that with write voltage of less than 2 V, the FeFET state can be free from the disturbance [28]. Without loss of generality, we simplify the array write operation by applying the same write pulse to all cells, so that each cell stores the same matching range for the search operation.

During the search phase, we sweep all the SLs associated with the array cells from 0 to 1 V, and plot the transient SA outputs versus different $V_{\text{SL}}$ values similar to Fig. 5(b). The search time of the array is determined according to the time required to maintain the predefined matching range stored in the cells. The results are demonstrated in Fig. 6(c)–(f). It is found that with the same search time, the FeCAM storage and search functionality are not significantly affected by the number of rows in the FeCAM array [Fig. 6(c) and (d)], which is reasonable as the parasitics associated with SLs (i.e., the FeFET gates) have negligible impact on the ML[2]. However, on the column line, the increasing number of columns equivalently adds additional discharge paths to the ML, inevitably affecting the discharge rate.

To ensure the matching range integrity as the number of columns increases, we propose to adapt the search time according to the different number of columns, which can be precharacterized at the design time. In this way, we can still keep the same matching bounds and the storage capability. Specifically, the associated capacitance of an ML grows linearly with the number of columns as in the following equation:

$$C_{\text{ML}} \approx C_{\text{pMOS}} + N \times (C_{\text{drain}} + C_{\text{parasitic}}) \quad (1)$$

where $C_{\text{ML}}$, $C_{\text{pMOS}}$, $C_{\text{drain}}$, and $C_{\text{parasitic}}$ are the associated capacitance of the ML, drain capacitance of the precharge pMOS, total drain capacitance of a FeCAM cell, and the parasitic capacitance of the interconnect for each cell, respectively. The values of $C_{\text{pMOS}}$ and $C_{\text{drain}}$ can be extracted from PTM 45-nm technology model [29], and the value of $C_{\text{parasitic}}$ is extracted from DESTINY [27]. $N$ is the number of columns in the array. The discharge time $\Delta t$ for the ML to drop by $\Delta V_{\text{ML}}$ ($\Delta V_{\text{ML}} = 0.5$ V in our setup as the search boundary) is described in the following equations:

$$\Delta t = C_{\text{ML}} \times \Delta V_{\text{ML}} / \left( \sum I_{\text{discharge}_i} \right)$$
$$= C_{\text{ML}} \times \Delta V_{\text{ML}} / (N \times \overline{I_{\text{discharge}}}) \quad (2)$$
$$\Delta t \approx \frac{\Delta V_{\text{ML}}}{I_{\text{discharge}}} \times (C_{\text{pMOS}}/N + C_{\text{drain}} + C_{\text{parasitic}}) \quad (3)$$

where $I_{\text{discharge}_i}$ is the $i$th single cell discharge current when the corresponding SL input $V_{SL_i}$ is at the boundary of the stored value range, and $\overline{I_{\text{discharge}}}$ denotes the average discharge current per cell from ML to ground during the search ($\overline{I_{\text{discharge}}}$ is around 25 nA per simulation). As the search time follows the same trends as the discharge time $\Delta t$, (3) indicates that the search time should be decreased as the number of columns increases in order to keep the upper and lower bounds of the matching range for the FeCAM array. We can use (3) to set the search time for a given number of columns to ensure the same matching bounds and storage capability. Fig. 6(e) and (f) summarizes the results for the matching bounds and the corresponding search time, respectively, which is consistent with the analysis above. It can be anticipated

---

[2]When the row number is much larger than 64, the impact of the parasitics associated with SLs is not negligible, thus demanding stronger SL drivers to distribute the input data across the array.

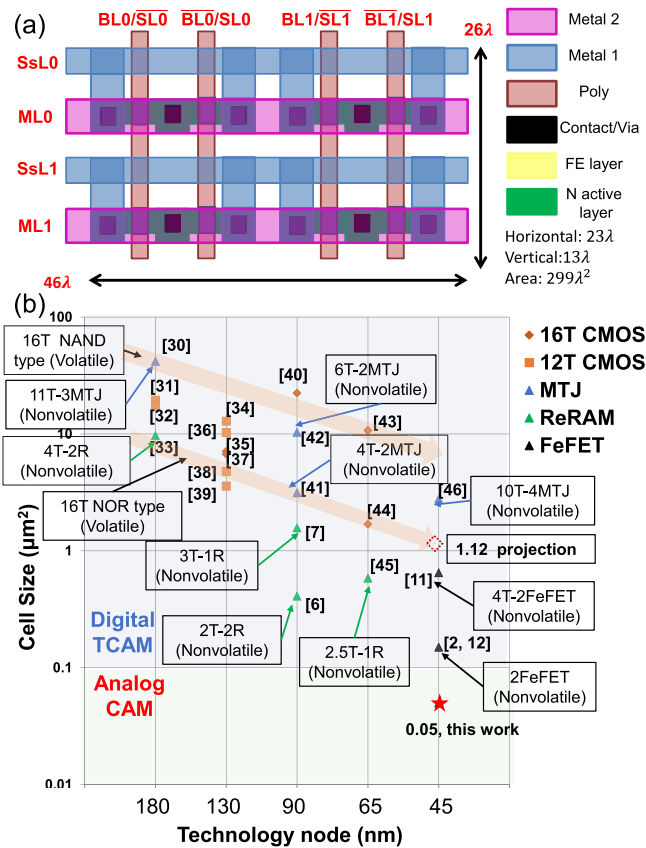

Fig. 7. (a) $2 \times 2$ FeCAM array layout. Note that $\lambda$ represents half-feature size $F$. (b) Comparisons of CAM cell area overhead per bit.

that as the number of columns increases, the search time will eventually reach a bound[3].

## IV. EVALUATION AND BENCHMARKING

In this section, we benchmark the performance of the FeCAM design in terms of area, search energy, search delay, and so on. We also present a straightforward application of FeCAM implementing a high-performance routing lookup table.

### A. FeCAM Cell

As discussed in Section III-B, the analog mode of FeCAM can encode multiple bits[4]. We sketch the layout of the $2 \times 2$ FeCAM array in Fig. 7(a), estimate the area per bit of FeCAM and compare the results with other existing TCAM work ([2], [6], [7], [11], [12], [30]–[46]) in Fig. 7(b). Since previous sections suggest that the analog mode of FeCAM is capable of storing continuous range for matching with eight discrete upper/lower bound levels in one cell, which is equivalent to the functionality of 3 TCAM cells, the area per bit of our FeCAM is 1/3 of that of the 2FeFET TCAM cell in [2], [12]. Fig. 7 shows that the area per bit of the analog mode of FeCAM is just 4.5% of that of a projected 16T CMOS TCAM design at a 45-nm technology node indicated by a red star in the chart. This area efficiency can enable compact CAM arrays, where fewer cells per row and fewer rows are required than a digital CAM design to store the same number of bits data. We present an IP router example as in Section IV-B.

---

[3]Strong precharge driver to charge the capacitance associated with the match line (ML) is demanded when the column number goes beyond 64, causing significant area and energy overhead.

[4]In this article, we use 3-bit/cell, as experimental proof of the 3-bit FeFET memory device has already been reported [24], further design and write scheme optimizations can enable more improvements in the memory density.

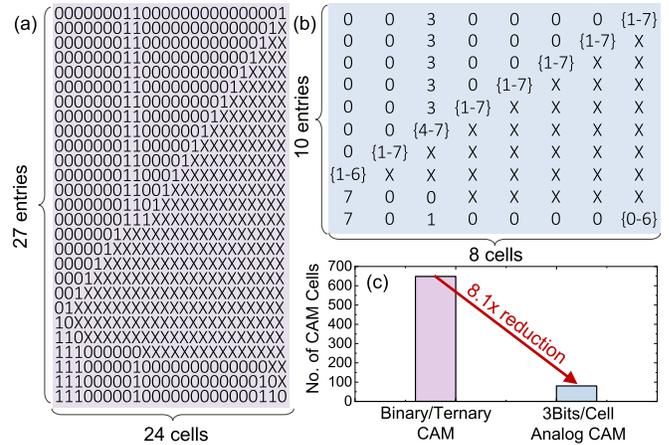

Fig. 8. Implementing a routing table covering a randomly chosen IP address range from 98 305 to 14 712 838 in the 24 bits IP address space with (a) binary/TCAM array. (b) 3 bits/cell analog FeCAM array. (c) Up to 8.1× reduction in the number of CAM cells can be achieved with FeCAM.

### B. FeCAM Application

We perform a thorough study on FeCAM and compare it with the conventional CMOS TCAM at array and application levels. We evaluate the search energy per bit for CMOS TCAM, digital and analog modes of FeCAM assuming 64-cell, 64- and 22-cell (one cell storing 3 bits) word sizes, respectively. We use an IP packet classification case to demonstrate the efficiency of FeCAM over other CAMs. While a typical CMOS 16T CAM array consumes 0.590 fJ/bit for the array search, the proposed FeCAM array can achieve 0.182 fJ/bit under digital mode (3.2× reduction with respect to CMOS TCAM) and 0.069 fJ/bit under analog mode for 3-bit search (8.6× reduction with respect to CMOS TCAM). Thus, the proposed FeCAM can provide not only stronger search capability and better memory density but also more energy saving. Note that the current design and simulations mainly validate the functionality of FeCAM analog mode concept, whereas FeCAM can be further optimized with respect to energy and performance.

Without loss of generality, a representative application of CAM array is demonstrated for IP address routing table in network devices (Fig. 8). As shown in Fig 8(a), a routing table with 24 bits routing prefix is considered, corresponding to 24 leading 1-bits in the subnet mask. A randomly chosen IP address range from 98 305 to 14 712 838 is then implemented in CMOS TCAM and FeCAM array, respectively, as shown in Fig. 8. To cover that range, 27 digital TCAM entries are necessary with 24 cells per entry; whereas only ten entries and eight cells per entry are required for a 3 bits/cell analog FeCAM array, indicating a considerable reduction (8.1×) of the array cells compared with the CMOS equivalent. Taking the cell area into consideration, the area and energy reduction of FeCAM-based routing table can be $60.5\times$ and $23.1\times$, respectively, compared with the CMOS TCAM-based routing table. These savings can be further improved by designing a 4 bits/cell analog FeCAM through device and design level optimization.

## V. DISCUSSIONS

In summary, compared with the prior CAM designs, the proposed FeCAM are featured with:

1) *Higher Flexibility:* With the same CAM cell structure, FeCAM can function as both digital and analog CAMs depending on the write and search schemes.

2) *Better Array Scalability:* The increasing size of the FeCAM array will not affect the functionality of both digital and analog CAM functionality, which is explained in Section III-C.

3) *Superior Energy Efficiency:* Due to the superior energy-efficient electric field-driven write mechanism of FeFETs, FeCAM can exhibit ultralow write energy consumption. On the other hand, the reduced number of CAM cell in analog FeCAM compared with its digital counterpart also reduces the search energy consumption.

4) *Higher Memory Density:* Section III-B shows that per FeFET controllable programming characteristics, a 3-bit data storage and search functionality can be achieved in the analog mode of FeCAM.

5) *Direct Analog Signal Processing Capability:* The analog mode of FeCAM enables novel computing functionality in analog domain, allowing the direct analog signal processing without analog to digital conversion, which may be very promising in IoT sensor scenarios.

This article represents an early exploratory device-circuit codesign of a universal FeCAM cell, which focuses on the demonstration of working principles of FeCAM cell and array. The nonidealities of FeFET technologies are not the focus of this article, but they are highly important for practical implementation of FeCAM. Given that $HfO_2$-based FeFET is still in its early development stage, several challenges still exist for this technology, especially for the analog/MLC states utilized in this article. The most important aspect is the degraded device-to-device variation for scaled FeFET [47]. Significant variation would limit the number of distinct upper/lower bounds that can be faithfully achieved in a FeCAM array. Improvement in this direction is still under intensive research. But several promising results have been demonstrated so far. For example, excellent cycle-to-cycle variation has been shown in a reasonably small ($W/L$=500 nm/500 nm) FeFET with 3 bits per cell [48]. For a large FeFET ($W/L$=10 $\mu$m/10 $\mu$m), even 3 bits per cell with well-controlled device-to-device variation has been demonstrated [24]. Another aspect is about the endurance of FeFET, which is limited to be around $10^5$ cycles [21]. But as discussed in [2] for digital FeCAM, the CAM applications, in general, may not require frequent write operations, as the search operations would likely be the most frequent, which is just FeFET read operations. Moreover, novel ferroelectric memory device structure is also actively pursued to bridge the endurance gap of FeFET and intrinsic ferroelectric [49].

Therefore, based on the discussions above, the benefits of the proposed FeCAM can be fully exploited with FeFET technology given the device improvement in the future.

## VI. CONCLUSION

In this article, we proposed a universal and compact CAM to support simultaneous digital and analog modes by exploiting the programmable analog/MLC states of FeFETs for the first time. A conceptual demonstration of analog CAM, as well as a FeFET-based CAM cell design have been demonstrated. Practical simulations show that the proposed FeCAM can encode multiple continuous ranges for matching with discrete upper/lower bound levels using just two FeFETs, thus significantly improving the memory density, area, and energy efficiency compared to the conventional 16T CMOS TCAM. This universal CAM design with both digital and analog search capabilities enables the compact memory array as well as flexible digital and analog signal processing in sensors, which is particularly critical for IoT applications.